\documentclass[acmsmall]{acmart}
\AtBeginDocument{%
  }

\usepackage{multirow}

\usepackage{microtype}
\usepackage{subfigure}
\usepackage{makecell}
\usepackage{color}
\usepackage{colortbl}

\usepackage{bm}
\usepackage{bbm}

\usepackage{cancel}

\usepackage[
  font = small,
  tableposition = top
]{caption}

\usepackage{algorithm}
\usepackage{algpseudocode}
\usepackage{booktabs}
\usepackage{multicol}
\usepackage{amsmath}
\usepackage{enumitem}

\usepackage{tikz}

\usepackage{tcolorbox}

\usepackage{threeparttable}
\usepackage[normalem]{ulem}

\setcopyright{acmlicensed}
\copyrightyear{2026}
\acmYear{2026}
\acmDOI{XXXXXXX.XXXXXXX}
\acmJournal{PACMSE}
\acmVolume{X}        
\acmNumber{ISSTA}    
\acmArticle{XX}      
\acmMonth{7}         

\acmISBN{978-1-4503-XXXX-X/2018/06}

\begin{document}

\title{HELO-APR: Enhancing Low-Resource Program Repair through Cross-Lingual Knowledge Transfer}

\author{Zhipeng Wang}
\email{kingwangzp@gmail.com}
\affiliation{
  \institution{State Key Laboratory for Novel Software Technology, Nanjing University}
  \city{Nanjing}
  \country{China}
}

\author{Boyang Yang}
\email{yby@ieee.org}
\affiliation{
  \institution{School of Artificial Intelligence (School of Software), Yanshan University}
  \city{Qinhuangdao}
  \country{China}
}

\author{Yidong Wan}
\email{522025320147@smail.nju.edu.cn}
\affiliation{
  \institution{State Key Laboratory for Novel Software Technology, Nanjing University}
  \city{Nanjing}
  \country{China}
}

\author{Liuye Guo}
\email{liuye.guo1@gmail.com}
\affiliation{
  \institution{State Key Laboratory for Novel Software Technology, Nanjing University}
  \city{Nanjing}
  \country{China}
}

\author{You Lv}
\email{2222452038@stmail.ujs.edu.cn}
\affiliation{
  \institution{Jiangsu University}
  \city{Zhenjiang}
  \country{China}
}

\author{Tao Zheng}
\email{zt@nju.edu.cn}
\affiliation{
  \institution{State Key Laboratory for Novel Software Technology, Nanjing University}
  \city{Nanjing}
  \country{China}
}

\author{Zhuowei Wang}
\authornote{Corresponding authors}
\email{zwwang@gdut.edu.cn}
\affiliation{
  \institution{School of Computer Science and Technology, Guangdong University of Technology}
  \city{Guangzhou}
  \country{China}
}

\author{Tieke He}
\authornotemark[1]
\email{hetieke@gmail.com}
\affiliation{
  \institution{State Key Laboratory for Novel Software Technology, Nanjing University}
  \city{Nanjing}
  \country{China}
}
\renewcommand{\shortauthors}{Wang et al.}

\begin{abstract}

Large Language Models (LLMs) perform well on automatic program repair (APR) for high-resource programming languages (HRPLs), but their effectiveness drops sharply in low-resource programming languages (LRPLs), due to a lack of sufficient verified buggy-fixed pairs for APR training.
To address this challenge, we propose \emph{HELO-APR} (\emph{H}igh-resource \emph{E}nabled \emph{LO}w-resource APR), a two-stage APR framework that enables cross-lingual transfer of repair knowledge from HRPLs to LRPLs.
\emph{HELO-APR} (1) constructs high-quality LRPL training data by synthesizing LRPL buggy-fixed pairs from HRPL counterparts, preserving defect type consistency while ensuring the synthesized code is idiomatic, and then (2) adopts a curriculum learning strategy that progressively performs HRPL repair learning, cross-lingual repair alignment, and LRPL repair adaptation, improving repair effectiveness in LRPLs.
Using C++ as the source HRPL and Ruby and Rust as the target LRPLs, experiments on xCodeEval show that \emph{HELO-APR} consistently outperforms strong baselines, increasing Pass@1 from 31.17\% to 48.65\% on DeepSeek-Coder-6.7B and from 1.67\% to 11.97\% on CodeLlama-7B, while improving syntactic validity by raising the average target compilation rate on CodeLlama from 49.77\% to 91.98\%.
On Defects4Ruby, \emph{HELO-APR} increases BLEU-4 from 61.20 to 66.79 and ROUGE-1 from 76.76 to 83.59 on CodeLlama-7B, indicating higher similarity to developer patches in real-world settings.
Finally, we conduct ablation studies to assess the necessity of each core component.
These results suggest that verified cross-lingual supervision provides a reusable approach for improving LLM-based repair in low-resource languages.

\end{abstract}

\maketitle

\section{Introduction}
\label{sec:introduction}

LLMs have advanced automated program repair (APR), but effectiveness still varies substantially across programming languages~\cite{xia2023automated,luo2025unlocking,yuan2022circle,zhang2024pydex,yang2025morepair,hossain2024deep, wang2025exploration,yang2024cref,luo2026fine}. LLMs perform well in high-resource programming languages (HRPLs), such as Python, Java, and C++~\cite{xia2022less,yuan2022circle,xia2024automated}, but remain weaker in low-resource programming languages (LRPLs), such as Ruby and Rust~\cite{baltajicross,wong2025investigating,wang2023towards}.
A primary driver is data scarcity: LRPLs lack sufficient validated buggy-fixed pairs to support learning robust repair patterns~\cite{zhang2023survey}.

To mitigate this gap, prior work has mainly focused on two directions. One direction constructs synthetic buggy-fixed data. For example, Wong et al.~\cite{wong2025investigating} construct buggy-fixed pairs by collecting LLM-generated buggy codes and synthesizing fixes with stronger LLMs. Although this pipeline scales, it is constrained by the generator's capabilities. Synthesized defects are often simplistic or repetitive (e.g., generic syntax errors) and fail to capture the semantic complexity of real-world bugs.

Another direction leverages high-resource data for transfer. Baltaji et al.~\cite{baltajicross} fine-tune LLMs on HRPL buggy-fixed pairs and apply them to LRPLs, motivated by the intuition that repair logic (e.g., for off-by-one errors) is shared across different programming languages. While such \textit{HRPL-only} training can yield modest improvements, its effectiveness remains limited. Our preliminary analysis (\S\ref{sec:rq1}) suggests that a key factor is \textit{syntactic interference}: when repairing target-language code, LLMs fine-tuned only on source languages sometimes generate source-language patches, which fail to compile under the target-language toolchain. For example, after fine-tuning CodeLlama on 10,000 C++ samples and evaluating on Ruby, 31.58\% of the top-1 patches are syntactically valid C++ rather than Ruby.

\begin{figure}[t]
    \centering
    \includegraphics[width=\linewidth]{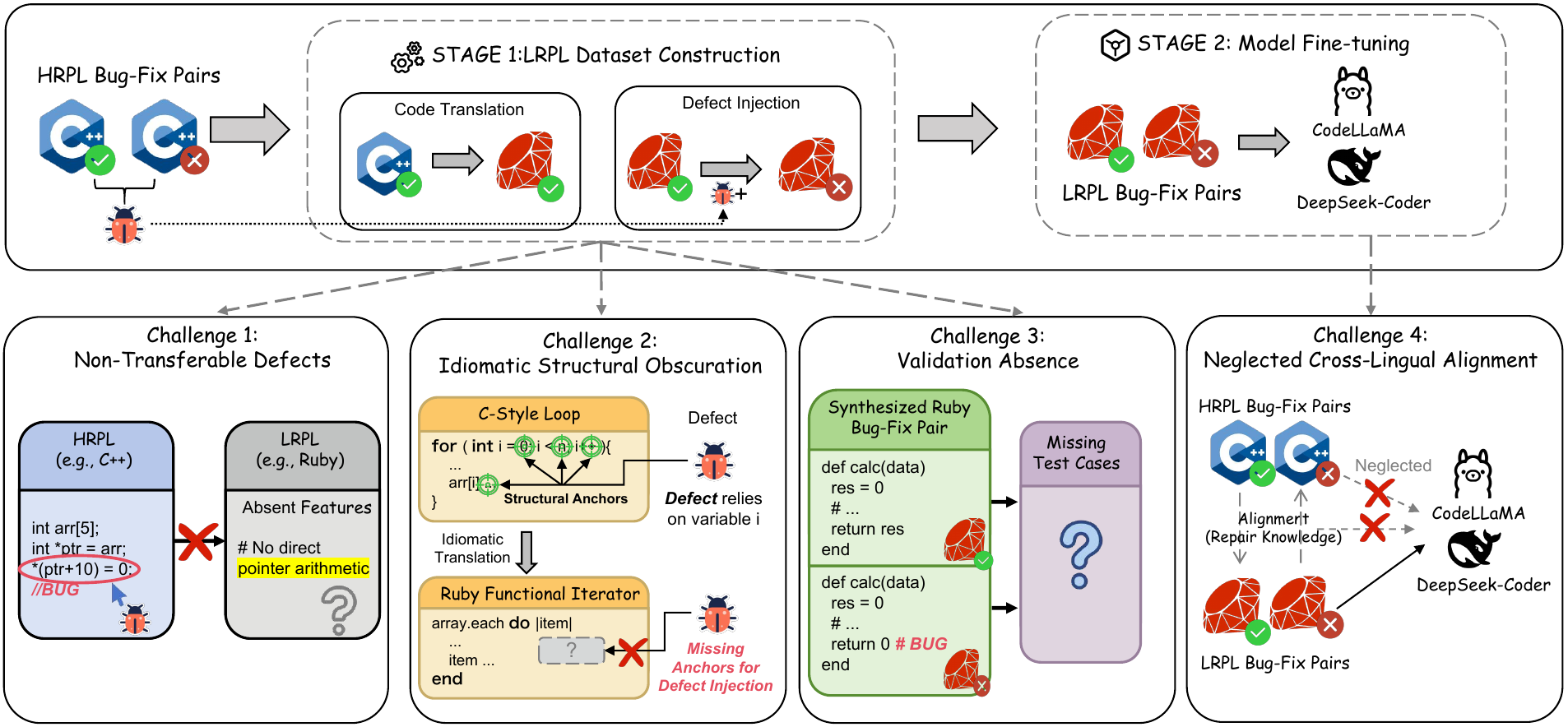}
    \caption{Overview of the cross-lingual APR workflow (upper) and four critical challenges limiting its effectiveness (lower): (1) \textit{Non-Transferable Defects}, (2) \textit{Idiomatic Structural Obscuration}, (3) \textit{Validation Absence}, and (4) \textit{Neglected Cross-Lingual Repair Alignment}.}
    \label{fig:trans_three_challenge}
\end{figure}

To bridge the gap between the lack of diversity in synthetic data and the syntactic interference introduced by direct transfer, we propose a third direction: \textit{translating HRPL buggy-fixed pairs into LRPLs and leveraging them for LLM fine-tuning} (as illustrated in Figure~\ref{fig:trans_three_challenge}). This approach combines the realistic complexity of HRPL data with the correct syntax of the target language. However, realizing this workflow is non-trivial. Specifically, during dataset construction, we encounter three challenges: (1) \textit{Non-Transferable Defects}, where certain source-language defects do not exist in the target language; (2) \textit{Idiomatic Structural Obscuration}, where abstract syntax patterns hide the structural anchors required for defect injection; and (3) \textit{Validation Absence}, which prevents verifying defect behavior due to missing test cases. In addition, training solely on synthetic data during fine-tuning is insufficient. As reflected in the fourth challenge, standard fine-tuning ignores the parallel relationship between HRPL and target LRPL data, leading to (4) \textit{Neglected Cross-Lingual Repair Alignment}.

To address these challenges, we propose HELO-APR, 
a two-stage approach for improving automated program repair in LRPLs.

First, for dataset construction, we synthesize high-fidelity LRPL 
buggy-fixed pairs. We apply \textit{Transferability Analysis} to pre-filter 
source samples with non-transferable defects, followed by a 
\textit{structure-constrained translation and injection strategy} that 
preserves defect injection anchors during translation, enabling reliable 
defect reproduction while maintaining idiomatic LRPL code. Finally, a 
\textit{test-driven verification} step is employed to select high-quality 
buggy-fixed pairs for subsequent training.

Second, for model fine-tuning, we address the challenge of 
\textit{Neglected Cross-Lingual Repair Alignment} via a 
\textit{curriculum learning-based strategy} that explicitly leverages the 
consistency of repair behaviors across parallel HRPL-LRPL data. 
Specifically, the model is trained across three progressive stages: 
\textit{HRPL repair learning}, \textit{cross-lingual repair alignment} 
(using synthesized parallel pairs to bridge linguistic gaps), and 
\textit{LRPL repair adaptation}, enabling effective knowledge transfer to 
LRPLs.

\textbf{Contributions.} This paper makes the following contributions:
\begin{itemize}
    \item \textbf{Data.} We construct a cross-lingual parallel repair dataset spanning one HRPL (C++) and two LRPLs (Ruby and Rust), maintaining defect consistency while ensuring idiomatic code in the target LRPLs. The released dataset is available at \url{https://github.com/ARROGANT666666/HELO-APR-ISSTA}.

    \item \textbf{Method.} We propose \textit{HELO-APR}, a two-stage approach that includes:
    (1) \textit{LRPL Dataset Construction}, which synthesizes high-fidelity LRPL buggy-fixed pairs; and
    (2) \textit{Cross-Lingual Knowledge Transfer}, which applies a three-stage curriculum to progressively transfer repair knowledge from HRPLs to LRPLs.

    \item \textbf{Evaluation.} We show that \textit{HELO-APR} substantially improves repair performance for LRPLs, increasing Pass@1 on DeepSeek-Coder-6.7B by 56.08\% relatively (from 31.17\% to 48.65\%) and over 7 times on CodeLlama-7B (from 1.67\% to 11.97\%) compared to zero-shot baselines.

    \item \textbf{Analysis.} We conduct ablation studies to isolate the impact of key components and evaluate generalization on a repository-level benchmark, Defects4Ruby.
\end{itemize}

\section{LRPL Dataset Construction}
\label{sec:lrpl_dataset_construction}

\begin{figure}
    \centering
    \includegraphics[width=0.9\linewidth]{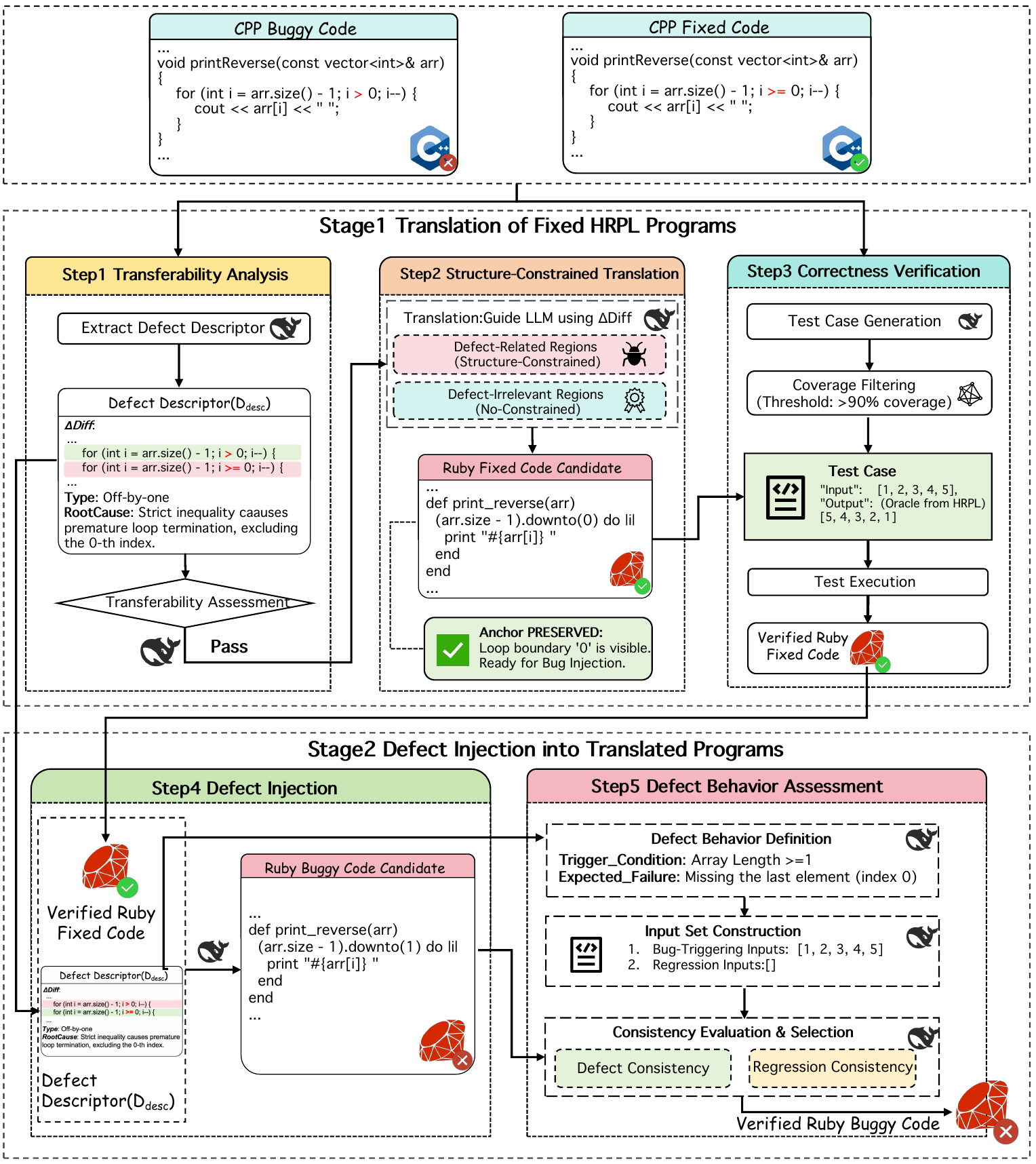}
    \caption{LRPL Dataset Construction.}
    \label{fig:framework}
\end{figure}

To alleviate the severe scarcity of high-quality buggy-fixed data in LRPLs, we propose an automated pipeline that synthesizes buggy-fixed pairs in LRPLs from HRPLs.
The overall pipeline consists of two high-level stages.
First, we translate fixed programs from HRPLs into LRPLs while preserving both functional correctness and code naturalness (\S\ref{sec:translation_fixed_hrpl}).
Second, we inject defects into the translated LRPL programs by referencing the corresponding defects in HRPLs, yielding LRPL buggy-fixed pairs that faithfully reproduce the defect behaviors exhibited in HRPLs (\S\ref{sec:defect_injection}).
Figure~\ref{fig:framework} provides an overview of the workflow.

\subsection{Translation of Fixed HRPL Programs}
\label{sec:translation_fixed_hrpl}

The goal of this stage is to map a fixed HRPL program $P_{hrpl}$ to a target LRPL program $P_{lrpl}$. 
Unlike general code translation tasks, this process requires not only \textit{functional correctness} but also the preservation of defect injection anchors necessary for subsequent injection. 
However, satisfying this dual requirement is challenging due to significant differences between languages. Taking HRPLs and typical LRPLs as examples, systemic differences in memory models, type systems, and control structures present two fundamental obstacles, corresponding to the first two challenges illustrated in Figure~\ref{fig:trans_three_challenge}:

\textbf{Non-transferable Defects.}
Certain HRPL defects are rooted in features absent from the target languages, such as pointer arithmetic, manual memory management, or undefined behaviors in C++.
Such defects cannot be reliably reproduced in LRPLs and must therefore be filtered out prior to the process.

\textbf{Idiomatic Structural Obscuration.} 
Even for transferable defects, syntactic idioms in LRPLs can obscure the structures required for defect injection. 
For instance, some LRPLs provide higher-level abstractions, such as Ruby’s \texttt{each} and \texttt{map} iterators, instead of explicit control-flow constructs. 
LLMs tend to adopt these idioms during translation to improve readability, which can inadvertently eliminate explicit loop boundaries, index variables, or branching conditions that serve as critical ``structural anchors'' for defect injection. 
For example, translating a C-style loop (e.g., \texttt{for (int i = 0; i < n; i++)}) into a functional iterator preserves functional behavior but compromises the structural fidelity required for accurate defect injection.

To address these challenges, we design a three-stage translation process:
\begin{enumerate}
    \item \textbf{Transferability Analysis}: filters out HRPL defects reliant on non-transferable language features (e.g., pointer arithmetic) to ensure reproducibility in the target LRPL.
    \item \textbf{Structure-Constrained Translation}: guides the LLM to preserve structural anchors required for defect injection within defect-relevant regions, thereby mitigating anchor loss caused by idiomatic abstraction.
    \item \textbf{Correctness Verification}: validates the target code using automatically generated test cases to guarantee functional correctness.
\end{enumerate}

\subsubsection{Transferability Analysis}

Prior to translation, it is essential to determine whether a specific defect in the source HRPL is reproducible in the target LRPL. To achieve this, we analyze the differences between the buggy version $P_{src}^{buggy}$ and the fixed version $P_{src}^{fixed}$ to extract key defect attributes.
As shown in Figure~\ref{fig:framework}, this process formalizes the defect into a concise descriptor:

\begin{equation}
    D_{desc} = \langle \textit{Type}, \textit{RootCause}, \Delta_{diff} \rangle
\end{equation}

Here, \textit{Type} and \textit{RootCause} represent the defect type and the underlying reason, respectively (e.g., ``Off-by-one'' and ``Incorrect relational operator `<='\,'').
Crucially, $\Delta_{diff}$ denotes the \textit{Patch}, capturing the exact code changes.

Based on this descriptor, we employ an LLM to assess the transferability of the defect. The LLM evaluates whether the defect characteristics defined in $\Delta_{diff}$ are transferable to the target LRPLs.
For instance, if $\Delta_{diff}$ contains pointer arithmetic or manual memory management specific to C++, the defect is deemed non-transferable and filtered out. 
Only defects rooted in language-agnostic logic proceed to the subsequent structure-constrained translation stage.

\subsubsection{Structure-Constrained Translation}
\label{sec:translation}
After identifying transferable cases, we translate the correct HRPL version $P_{src}^{fixed}$ into the target LRPL program $P_{tgt}^{fixed}$. 
Unlike traditional code translation, this task requires not only functional equivalence but also the preservation of structural anchors that are essential for subsequent defect injection. 
Consequently, we formulate the translation process as a structure-constrained translation task.

Specifically, we employ an LLM to generate the target LRPL program.
To ensure that the generated code supports subsequent defect injection, we utilize the \textit{patch $\Delta_{diff}$} within $D_{desc}$ to distinguish between regions that require structural preservation and those where idiomatic adaptation is permitted.
Accordingly, we implement two types of constraints based on the regions involved:

\begin{itemize}
    \item \textbf{Defect-Related Regions ($\Delta_{diff}$ Scope):}
    For code segments within $\Delta_{diff}$, we prioritize preserving structural anchors required for defect injection over idiomatic rewriting. For example, when the fix involves a boundary condition (e.g., \texttt{i < n}), we preserve the corresponding defect-controlling anchor. Refactoring into higher-level abstract iteration constructs is permitted only when the defect-controlling anchor can be explicitly retained; otherwise, such refactoring is avoided to prevent anchor loss.

    \item \textbf{Defect-Irrelevant Regions (Complement of $\Delta_{diff}$):} 
    For code segments outside $\Delta_{diff}$, we allow idiomatic rewriting without structural constraints,
    since these parts do not trigger the defect.
    This improves the naturalness and readability of the translated code.
\end{itemize}

This strategy ensures that the generated $P_{tgt}^{fixed}$ retains the structural anchors essential for defect injection while maintaining overall code naturalness.
Subsequently, we employ a test-driven verification process (\S\ref{sec:verification}) to screen for functionally correct translation candidates.

\subsubsection{Correctness Verification}
\label{sec:verification}
In this subsection, we verify the generated target program through a two-step process: 1) automated test generation, and 2) translation validation. This ensures the target program behaves consistently with the HRPL version.

\noindent \textbf{(1) Automated Test Generation.} 
Leveraging the correct HRPL program $P_{src}^{fixed}$, we use a test generator $G_{test}$ to construct
a test suite $S_{test}$. The goal is to cover defect-relevant execution paths and mitigate the risk of
``pseudo-equivalence'' caused by insufficient testing. The workflow consists of two steps:

\begin{enumerate}
    \item \textit{Test Case Generation:} We prompt an LLM to generate diverse inputs based on the program structure.
    For each input, we execute $P_{src}^{fixed}$ and record the program outputs, forming language-agnostic input-output pairs.
   \item \textit{Quality Filtering:} Generated test cases are filtered using coverage-based criteria.
Following Cassano et al.~\cite{cassano2024knowledge}, we require both line and branch coverage, measured by \textsc{gcov} (GCC), to exceed a threshold $\tau$ (set to 90\%); test cases failing to meet this requirement are discarded.

    \begin{equation}
        S_{test} \subseteq G_{test}(P_{src}^{fixed}),
        \quad
        \text{Cov}_{line}(S_{test}) \ge \tau \;\land\; \text{Cov}_{branch}(S_{test}) \ge \tau .
    \end{equation}
\end{enumerate}

The filtered suite $S_{test}$ serves as the reference oracle for subsequent verification.

\noindent \textbf{(2) Translation Validation.}
We validate the translated program using $S_{test}$.
To improve robustness, we perform up to $m$ translation attempts.
In each attempt $j$, the LLM produces a candidate program $P_{cand}^{(j)}$, which is executed on $S_{test}$.
We select the first candidate that passes all tests:
\begin{equation}
    j^\star = \min \{\, j \in \{1,\dots,m\} \mid Pass(P_{cand}^{(j)}, S_{test}) \,\},
    \quad
    P_{tgt}^{fixed} = P_{cand}^{(j^\star)} .
\end{equation}

\subsection{Defect Injection into Translated Programs}
\label{sec:defect_injection}

Given the verified target program $P_{tgt}^{fixed}$, our objective is to construct a buggy counterpart
$P_{tgt}^{buggy}$ whose observable behavior matches the defect behavior exhibited by the original HRPL program.
Rather than replicating syntactic edits, we aim to preserve the defect's behavioral manifestation in the LRPL.
To this end, we generate defect candidates guided by the defect descriptor $D_{desc}$ and evaluate them
using defect behavior assessment.

\subsubsection{Defect Injection}

Guided by the defect type ($Type$), root cause ($RootCause$), and the patch ($\Delta_{diff}$) contained in $D_{desc}$, we inject defects into the target program $P_{tgt}^{fixed}$.
Since the same defect logic can be implemented in multiple ways in the target language, we generate a set of candidate buggy programs:
\begin{equation}
    \mathcal{C} = \{P_1, P_2, \dots, P_n\}
\end{equation}
Each candidate $P_i \in \mathcal{C}$ aims to reproduce the original defect behavior, while its correctness is determined through subsequent validation.

\subsubsection{Defect Behavior Assessment}
\label{sec:defect_validation}

After generating the candidate set $\mathcal{C}$, we select the candidate that best
reproduces the defect behavior of the original HRPL program while minimizing unintended
regressions, evaluated along two complementary dimensions: defect consistency and
regression consistency.

\noindent \textbf{(1) Defect Behavior Definition.}
We characterize each defect using a behavioral specification derived from the defect
descriptor $D_{desc}$.
Specifically, an LLM generates a structured defect behavior description:
\begin{equation}
    B_{\text{defect}} = \langle \textit{TriggerCondition},\; \textit{ExpectedFailure} \rangle .
\end{equation}
Here, \textit{TriggerCondition} specifies the input conditions under which the defect is
activated, and \textit{ExpectedFailure} describes the expected observable failure pattern
(e.g., crash, exception, or incorrect output).

\noindent \textbf{(2) Input Set Construction.}
Guided by $B_{\text{defect}}$, we use an LLM to generate test inputs that are likely to
trigger or not trigger the defect, and execute each input on both the buggy and fixed
HRPL programs to compare their observable behaviors.
We then construct two disjoint input sets:
\begin{itemize}
    \item \textit{Bug-Triggering Inputs ($S_{\text{trigger}}$):}
    inputs $t$ on which the buggy and fixed HRPL programs exhibit different observable behaviors,
    \begin{equation}
        S_{\text{trigger}} = \{\, t \mid P_{src}^{buggy}(t) \neq P_{src}^{fixed}(t) \,\},
    \end{equation}
    indicating that the defect is activated under these inputs.
    
    \item \textit{Regression Inputs ($S_{\text{reg}}$):}
    inputs $t$ on which both versions exhibit identical observable behaviors,
    \begin{equation}
        S_{\text{reg}} = \{\, t \mid P_{src}^{buggy}(t) = P_{src}^{fixed}(t) \,\}.
    \end{equation}
\end{itemize}
Bug-triggering inputs are essential for defect reproduction; if no input in $S_{\text{trigger}}$ can be generated, the instance is discarded.
Regression inputs are optional and may not exist for all instances.

\noindent \textbf{(3) Consistency Evaluation.}
For each candidate program $P_i \in \mathcal{C}$, we evaluate its behavior on the constructed input sets:
\begin{itemize}
    \item \textit{Defect Consistency.}
    On bug-triggering inputs, the candidate is expected to exhibit a failure behavior
    consistent with \textit{ExpectedFailure}.
    Since observable failure manifestations may differ across programming languages,
    we assess consistency using a behavioral equivalence relation rather than strict output equality.
    Specifically, two executions are considered equivalent if they exhibit the same failure category
     (e.g., crash, exception, or incorrect output).
    The defect consistency score is defined as:
    \begin{equation}
        N_{\text{defect}}(P_i) =
        \sum_{t \in S_{\text{trigger}}}
        \mathbf{1}\big(P_i(t) \equiv P_{src}^{buggy}(t)\big),
    \end{equation}
    where $\equiv$ denotes behavioral equivalence under the above failure criteria.
    
    \item \textit{Regression Consistency.}
    On regression inputs, the candidate is expected to preserve the correct behavior
    of the fixed target program $P_{tgt}^{fixed}$.
    We quantify regression consistency as:
    \begin{equation}
        N_{\text{reg}}(P_i) =
        \sum_{t \in S_{\text{reg}}}
        \mathbf{1}\big(P_i(t) = P_{tgt}^{fixed}(t)\big).
    \end{equation}
\end{itemize}

\noindent \textbf{(4) Final Selection.}
Since only candidates that successfully reproduce the defect behavior are meaningful,
we rank candidates using a two-level criterion that prioritizes defect consistency
and uses regression consistency as a secondary criterion.
Formally, the final buggy program is selected as:
\begin{equation}
    P_{tgt}^{buggy}
    =
    \arg\max_{P_i \in \mathcal{C}}
    \big\langle N_{\text{defect}}(P_i),\; N_{\text{reg}}(P_i) \big\rangle .
\end{equation}

\section{Cross-Lingual Knowledge Transfer}
\label{sec:knowledge_transfer}

Although the dataset construction framework in \S\ref{sec:lrpl_dataset_construction}
mitigates the scarcity of high-quality LRPL buggy-fixed pairs, constructing verified LRPL training data remains costly. Under such strict data constraints, effectively transferring repair knowledge from HRPLs to LRPLs becomes critical.
A naive solution is to directly fine-tune an LLM on the synthesized LRPL dataset.
However, this approach is suboptimal, as it fails to exploit the semantic alignment between HRPL and LRPL programs and therefore underutilizes the rich repair supervision available in HRPLs.

\begin{figure}
    \centering
    \includegraphics[width=1\linewidth]{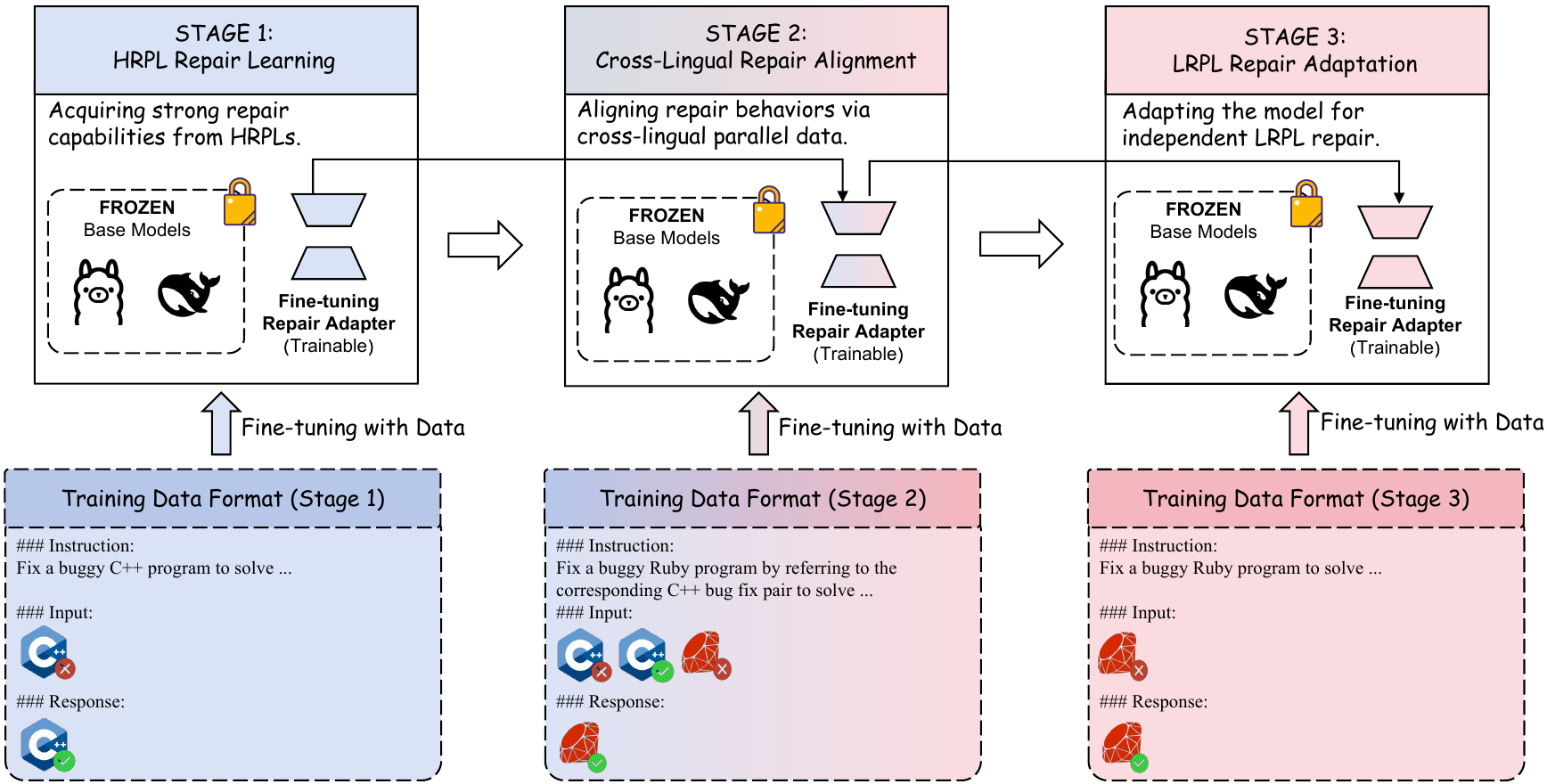}
    \caption{Three-stage curriculum learning framework of HELO-APR. The model is progressively trained via HRPL repair learning, cross-lingual repair alignment using parallel data, and LRPL repair adaptation. The backbone is frozen throughout training, and only repair adapters are fine-tuned to enable efficient and stable knowledge transfer.}
    \label{fig:finetune}
\end{figure}

To address this limitation, we propose a \textit{three-stage curriculum learning} framework
that progressively transfers repair knowledge from HRPLs to LRPLs. The key intuition is to first establish robust repair reasoning in a high-resource setting,
then explicitly align repair behaviors across languages using parallel buggy-fixed pairs,
and finally adapt the model to independently perform LRPL repair. By decoupling semantic repair logic from language-specific syntax, this curriculum enables
effective and stable cross-lingual knowledge transfer.
The overall workflow is illustrated in Figure~\ref{fig:finetune}.


\paragraph{\textbf{Problem Formulation}.}
Let $\mathcal{D}_{src}=\{(P_{src}^{buggy},P_{src}^{fixed})\}$ denote the HRPL buggy-fixed dataset, and let $\mathcal{D}_{tgt}=\{(P_{tgt}^{buggy},P_{tgt}^{fixed})\}$ denote the synthesized LRPL dataset.
For instances where both HRPL and LRPL pairs are successfully constructed and verified, we form a parallel dataset
$\mathcal{D}_{para}=\{(P_{src}^{buggy},P_{src}^{fixed},P_{tgt}^{buggy},P_{tgt}^{fixed})\}$.

Let $\theta$ denote the trainable parameters of the repair adapters, while the backbone
parameters are frozen throughout training.
Training proceeds sequentially: parameters learned in Stage $k$ are used to initialize
Stage $k{+}1$.
Each stage minimizes a negative log-likelihood (NLL) objective defined on its corresponding dataset.

\noindent \textbf{Stage 1: HRPL Repair Learning.}
In the first stage, we train the repair adapters on HRPL repair data to learn strong
repair reasoning in a high-resource setting.
Specifically, the model is trained to generate the fixed program given the buggy HRPL program
(\textit{Input:} $P_{src}^{buggy}$ $\rightarrow$ \textit{Output:} $P_{src}^{fixed}$).
This stage equips the model with general repair capabilities for complex semantic defects.
However, due to strong coupling with HRPL-specific syntax, these capabilities exhibit
limited transferability to LRPLs, motivating the subsequent alignment stage.
The training objective is:
\begin{equation}
    \mathcal{L}_{\text{init}}(\theta)
    = -\mathbb{E}_{(P_{src}^{buggy}, P_{src}^{fixed}) \sim \mathcal{D}_{src}}
    \big[\log P_\theta(P_{src}^{fixed} \mid P_{src}^{buggy})\big].
\end{equation}

\noindent \textbf{Stage 2: Cross-Lingual Repair Alignment.}
The second stage explicitly aligns repair behaviors across languages using the parallel
dataset $\mathcal{D}_{para}$.
We adopt a demonstration-based fine-tuning strategy, in which the HRPL buggy-fixed pair
serves as a repair demonstration for guiding LRPL repair.
Concretely, the model is conditioned on the HRPL buggy code, its corresponding fix,
and the LRPL buggy code, and is trained to generate the LRPL fix
(\textit{Input:} $P_{src}^{buggy}, P_{src}^{fixed}, P_{tgt}^{buggy}$
$\rightarrow$ \textit{Output:} $P_{tgt}^{fixed}$).
This formulation encourages the transfer of repair knowledge while preventing direct copying
of source-language syntax.
The alignment objective is defined as:
\begin{equation}
    \mathcal{L}_{\text{align}}(\theta)
    = -\mathbb{E}_{(P_{src}^{buggy}, P_{src}^{fixed}, P_{tgt}^{buggy}, P_{tgt}^{fixed}) \sim \mathcal{D}_{para}}
    \big[\log P_\theta(P_{tgt}^{fixed}
    \mid P_{src}^{buggy}, P_{src}^{fixed}, P_{tgt}^{buggy})\big].
\end{equation}

\noindent \textbf{Stage 3: LRPL Repair Adaptation.}
In the final stage, we adapt the model to the LRPL repair setting by fine-tuning the repair
adapters exclusively on LRPL data, without any HRPL demonstrations.
The model is trained to independently generate the fixed LRPL program given the buggy LRPL
program (\textit{Input:} $P_{tgt}^{buggy}$ $\rightarrow$ \textit{Output:} $P_{tgt}^{fixed}$).
This stage removes the model's reliance on source-language context and enables direct
application to LRPL repair tasks.
The adaptation objective is:
\begin{equation}
    \mathcal{L}_{\text{adapt}}(\theta)
    = -\mathbb{E}_{(P_{tgt}^{buggy}, P_{tgt}^{fixed}) \sim \mathcal{D}_{tgt}}
    \big[\log P_\theta(P_{tgt}^{fixed} \mid P_{tgt}^{buggy})\big].
\end{equation}

\section{Evaluation}
\label{sec:evaluation}

\subsection{Research Questions}
\label{sec:rq}
We evaluate \textit{HELO-APR} by answering the following research questions:

\begin{description}
    \item[RQ1 (Effectiveness):]
    How effective is \textit{HELO-APR} compared to existing baselines on low-resource program repair benchmarks?

    \item[RQ2 (Dataset Construction):]
    How do different components of the proposed dataset construction pipeline
    (i.e., Transferability Analysis, Structure-Constrained Translation \& Injection,
    and Test-Driven Verification)
    contribute to the quality of the synthesized LRPL dataset?

    \item[RQ3 (Cross-Lingual Knowledge Transfer):]
    How does each stage in the proposed curriculum learning framework contribute to the final repair performance?

    \item[RQ4 (Generalizability):]
    How well does \textit{HELO-APR} generalize to real-world software development scenarios?
\end{description}

\subsection{Dataset}
\label{sec:dataset}

\paragraph{Dataset Overview.}
Our experiments involve three data sources serving distinct roles in the proposed approach.
Specifically, we use
(i) HRPL data from \textsc{xCodeEval} for synthesizing LRPL buggy-fixed training pairs,
(ii) LRPL benchmarks from \textsc{xCodeEval} for controlled evaluation under low-resource settings, and
(iii) a real-world LRPL dataset, \textsc{Defects4Ruby}~\cite{dehghandefects4ruby}, for external validation in practical software development scenarios.

\paragraph{HRPL Source Data (\textsc{xCodeEval})~\cite{khan2023xcodeeval}.}
We select \textit{C++} as the HRPL due to the availability of abundant high-quality submissions.
From the APR task of \textsc{xCodeEval}, we randomly sample 10,000 C++ buggy-fixed pairs $(P^{buggy}_{src}, P^{fixed}_{src})$ from the training split.
These pairs are used to translate $P^{fixed}_{src}$ into LRPL and inject the defect derived from $(P^{buggy}_{src}, P^{fixed}_{src})$ to synthesize LRPL buggy-fixed training pairs (\S\ref{sec:lrpl_dataset_construction}).

\paragraph{Controlled LRPL Benchmark (\textsc{xCodeEval}).}
To evaluate repair performance under controlled low-resource settings, we adopt the \textit{Compact Set} of \textsc{xCodeEval} for two LRPLs: \textit{Ruby} and \textit{Rust}.
This benchmark enables systematic comparison across methods under consistent problem settings.

\paragraph{Real-World Evaluation Dataset (\textsc{Defects4Ruby}).}
To assess the generalizability of \textsc{HELO-APR} beyond contest-style benchmarks,
we further conduct experiments on \textsc{Defects4Ruby}~\cite{dehghandefects4ruby}, a large-scale collection of
real-world Ruby buggy-fixed pairs mined from GitHub projects.
This dataset is used solely for external validation.

\vspace{-5pt}

\subsection{Baselines}
\label{sec:baselines}
To evaluate the effectiveness of \textsc{HELO-APR}, we compare it against four baseline methods:

\noindent \textit{(1) Zero-Shot Learning:} The base LLM is prompted to repair buggy code directly without any parameter updates. This baseline reflects the model's inherent capability on LRPLs\cite{pourpanah2022review}.

\noindent \textit{(2) HRPLs-Only:} The model is fine-tuned solely on the HRPL dataset and then evaluated on LRPL benchmarks. This baseline captures the performance achievable without leveraging any synthesized LRPL data\cite{cassano2024knowledge}.

\noindent \textit{(3) Wong et al.\cite{wong2025investigating}} : LRPL buggy code is first generated by the base model based on problem descriptions from the source data (HRPL). \textit{DeepSeek-V3.2-Chat} is then employed to synthesize the corresponding repair rationales and fixed code. The model is subsequently fine-tuned on these verified  buggy-fixed pairs.

\noindent \textit{(4) LANTERN~\cite{luo2025unlocking}:} LANTERN is a multi-agent framework for cross-lingual program repair. For buggy LRPL programs that cannot be directly repaired, it translates them into a suitable HRPL selected by an LLM-based analyzer, performs repair in the target language, and back-translates successful fixes. Failed attempts are iteratively refined based on previous outputs. Unlike \textsc{HELO-APR}, it requires no additional training but incurs higher inference cost due to its multi-step pipeline. 
For a fair comparison, each iteration in LANTERN is treated as an independent inference step, ensuring a comparable inference budget with HELO-APR.

\subsection{Evaluation Metrics}
\label{sec:metric}

We employ two metrics to evaluate both the functional correctness and the syntactic validity of the generated patches.

\noindent\textbf{Pass@k.}
Following standard practice~\cite{luo2025unlocking}, we use Pass@k as the primary metric to evaluate repair performance.
Pass@k measures the probability that at least one of the top-$k$ generated patches passes all unit tests, and is computed using the unbiased estimator:
\begin{equation}
\text{Pass@k} := \underset{\text{Problems}}{\mathbb{E}} \left[ 1 - \frac{\binom{n-c}{k}}{\binom{n}{k}} \right],
\end{equation}
where $n$ is the number of generated candidates per problem, $c$ is the number of test-passing candidates under the \textsc{ExecEval} environment, and $k$ denotes the candidate budget.
In our experiments, we set $n=5$ and report \textit{Pass@1}, \textit{Pass@3}, and \textit{Pass@5}.

\noindent\textbf{Compilation Rate (CR).}
We use the compilation rate to assess the syntactic correctness of generated patches.
It measures the percentage of \emph{top-1} generated patches that compile (for Rust) or parse
successfully (for Ruby) in the \textsc{ExecEval} environment\cite{khan2023xcodeeval}, independent of their functional correctness.
\vspace{-5pt}

\subsection{Implementation Details}
\label{sec:implementation}

\smallskip
\noindent\textit{Model Selection.}
For dataset construction (\S\ref{sec:lrpl_dataset_construction}),
\textit{DeepSeek-V3.2-Chat} is used for translation and defect injection.
Specifically, we generate $m=5$ candidates for structure-constrained translation, and similarly generate $n=5$ candidates for defect injection. 
For cross-lingual knowledge transfer (\S\ref{sec:knowledge_transfer}), we adopt two backbone models:
\textit{CodeLlama-7B-Instruct} and \textit{DeepSeek-Coder-6.7B-Instruct}.

\smallskip
\noindent\textit{Training Configuration.}
Models are fine-tuned with \textit{QLoRA}, employing 4-bit NF4 quantization with double quantization enabled.
The backbone parameters are frozen, and only LoRA adapters are fine-tuned.
The LoRA rank is set to $r=8$ with $\alpha=16$.
Fine-tuning uses a global batch size of 32 and a maximum sequence length of 8192 tokens.
The AdamW optimizer is applied with a learning rate of $1\times10^{-4}$ and a learning rate schedule with 30 warmup steps.

\smallskip
\noindent\textit{Inference and Evaluation.}
All evaluations use a temperature of $T=1.0$.
For \textit{Pass@1}, we report the top-1 sample under this setting,
while for \textit{Pass@k} ($k>1$), multiple samples are drawn to estimate performance under larger candidate budgets.
All generated patches are executed within the isolated \textsc{ExecEval} sandbox.

\smallskip
\noindent\textit{Infrastructure.}
All experiments are conducted on a cluster equipped with 8 $\times$ NVIDIA A100 (40GB) GPUs.

\section{Evaluation Results}
\subsection{RQ1: Effectiveness}
\label{sec:rq1}

To evaluate the overall effectiveness of \textsc{HELO-APR}, we compare it against four representative baselines. We report \textit{Pass@k} ($k \in \{1, 3, 5\}$) to assess repair success under different candidate budgets, together with target compilation rate (CR$_T$) and source-language compilation rate (CR$_S$) to quantify syntactic validity and cross-lingual syntactic interference. Table~\ref{tab:main_results} presents the detailed results.

\begin{table*}[t]
\centering
\caption{Main Results on xCodeEval. We report \textit{Pass@k} (k=1, 3, 5), \textit{Target Compilation Rate (CR$_T$)}, and \textit{Source/C++ Compilation Rate (CR$_S$)}. High CR$_S$ indicates syntactic interference. Best results (for performance metrics) in each backbone are highlighted in \textbf{bold}.}
\label{tab:main_results}
\renewcommand{\arraystretch}{1.15}
\setlength{\tabcolsep}{2.5pt} 
\begin{footnotesize}
\begin{tabular}{l ccccc @{\hspace{1em}} ccccc @{\hspace{1em}} ccccc}
\toprule
\multirow{2}{*}{\textbf{Method}} 
& \multicolumn{5}{c}{\textbf{Ruby}} 
& \multicolumn{5}{c}{\textbf{Rust}} 
& \multicolumn{5}{c}{\textbf{Average}} \\
\cmidrule(lr){2-6} \cmidrule(lr){7-11} \cmidrule(lr){12-16}
& P@1 & P@3 & P@5 & CR$_T$ & CR$_S$ 
& P@1 & P@3 & P@5 & CR$_T$ & CR$_S$ 
& P@1 & P@3 & P@5 & CR$_T$ & CR$_S$ \\
\midrule
\multicolumn{16}{l}{\textit{\textbf{Backbone A: CodeLlama-7B-Instruct}}} \\
\midrule
\textit{Zero-Shot}      
& 2.17 & 4.33 & 7.12 & 71.21 & 0.00      
& 1.16 & 3.47 & 4.62 & 28.32 & 0.00      
& 1.67 & 3.90 & 5.87 & 49.77 & 0.00 \\
\textit{HRPLs-Only}     
& 5.88 & 9.60 & 11.46 & 61.30 & 31.58      
& 2.31 & 4.05 & 5.20 & 29.48 & 0.00      
& 4.10 & 6.83 & 8.33 & 45.39 & 15.79 \\
\textit{Wong et al.}      
& 6.19 & 10.53 & 12.69 & 89.78 & 0.00       
& 4.05 & 7.51 & 9.25 & 49.71 & 0.00       
& 5.12 & 9.02 & 10.97 & 69.75 & 0.00 \\
\textit{LANTERN}
& 3.55 & 4.00 & 5.60 & \textbf{98.66} & 0.00
& 3.79 & 5.80 & 9.20 & 71.56 & 0.00
& 3.67 & 4.90 & 7.40 & 85.11 & 0.00 \\
\rowcolor{gray!10}
\textsc{HELO-APR} 
& \textbf{12.38} & \textbf{17.34} & \textbf{19.50} & 97.83 & 0.00 
& \textbf{11.56} & \textbf{18.50} & \textbf{19.65} & \textbf{86.13} & 0.00  
& \textbf{11.97} & \textbf{17.92} & \textbf{19.58} & \textbf{91.98} & 0.00 \\
\midrule
\multicolumn{16}{l}{\textit{\textbf{Backbone B: DeepSeek-Coder-6.7B-Instruct}}} \\
\midrule
\textit{Zero-Shot}      
& 24.77 & 38.08 & 43.65 & 92.88 & 0.00    
& 37.57 & 49.71 & 54.91 & 74.57 & 0.00    
& 31.17 & 43.90 & 49.28 & 83.73 & 0.00 \\
\textit{HRPLs-Only}     
& 27.86 & 34.37 & 36.53 & 91.33 & 4.64    
& 42.77 & 50.87 & 52.02 & 77.46 & 0.00    
& 35.32 & 42.62 & 44.28 & 84.40 & 2.32 \\
\textit{Wong et al.}      
& 33.75 & 41.80 & 46.44 & 97.52 & 0.00       
& 45.09 & 52.02 & 56.65 & 82.66 & 0.00    
& 39.42 & 46.91 & 51.55 & 90.09 & 0.00 \\
\textit{LANTERN}
& 30.56 & 43.00 & 50.50 & \textbf{100.00} & 0.00
& 38.15 & 56.64 & \textbf{63.01} & 76.90 & 0.00
& 34.36 & 49.82 & 56.75 & 88.45 & 0.00 \\

\rowcolor{gray!10}
\textsc{HELO-APR} 
& \textbf{41.80} & \textbf{50.15} & \textbf{53.56} & 98.45 & 0.00    
& \textbf{55.49} & \textbf{60.12} & 62.43 & \textbf{87.28} & 0.00    
& \textbf{48.65} & \textbf{55.14} & \textbf{58.00} & \textbf{92.87} & 0.00 \\
\bottomrule
\end{tabular}
\end{footnotesize}
\end{table*}

The results demonstrate the superiority of \textsc{HELO-APR} in three key aspects:

\smallskip
\noindent\textbf{(1) Superior Overall Effectiveness.}
As evidenced in Table~\ref{tab:main_results}, \textsc{HELO-APR} achieves the best average Pass@k results and outperforms baselines  across most experimental settings.
In terms of functional correctness (\textit{Pass@k}), \textsc{HELO-APR} achieves substantial gains: on the CodeLlama backbone, it more than doubles the \textit{Pass@1} of the strongest baseline, \textit{Wong et al.} (11.97\% vs. 5.12\%), and maintains a clear lead on DeepSeek-Coder (48.65\% vs. 39.42
This superiority extends to compilation rate (\textit{$CR_T$}), where \textsc{HELO-APR} achieves near-perfect compilation rates (e.g., avg. 91.98\% on CodeLlama), competitive with or surpassing all other baselines and demonstrating robust repair capabilities under strict constraints.

\smallskip
\noindent\textbf{(2) Effective Mitigation of Cross-Lingual Syntactic Interference.}
Table~\ref{tab:main_results} shows that \textit{HRPLs-Only} suffers from syntax interference, while \textsc{HELO-APR} effectively mitigates this problem.
First, \textit{HRPLs-Only} fails to stop the source syntax from leaking.
It has a high Source Compilation Rate ($CR_S$) of 31.58\%, which means nearly one-third of the generated patches are actually C++ code.
Correspondingly, its Target Compilation Rate ($CR_T$) drops to only 61.30\%.
In contrast, using the same source data, \textsc{HELO-APR} reduces $CR_S$ to 0.00\% and raises $CR_T$ to 97.83\%, competitive with or surpassing all other baselines.
This indicates that \textsc{HELO-APR} fixes the interference issue: it learns the repair logic but strictly follows the target language rules.

\smallskip
\noindent\textbf{(3) Importance of High-Resource Reference Data.}
Comparing \textsc{HELO-APR} with \textit{Wong et al.} highlights the value of using C++ (HRPL) data as a reference.
\textit{Wong et al.} relies on LLMs to generate buggy-fixed pairs from scratch, lacking the guidance of existing HRPL buggy-fixed pairs.
In contrast, \textsc{HELO-APR} uses high-quality C++ pairs as a reference to synthesize the corresponding LRPL training data.
This guidance leads to superior performance.
For example, on the Ruby task (CodeLlama), \textsc{HELO-APR} achieves Pass@1/3/5 scores of 12.38\%, 17.34\%, and 19.50\%, consistently outperforming \textit{Wong et al.} (6.19\%, 10.53\%, and 12.69\%).
This comparison highlights the benefits of using HRPL buggy-fixed pairs as references: they provide stronger supervision than generating pairs from scratch and improve LRPL repair.

\smallskip
\noindent\textbf{(4) Advantage over Training-Free Translation Pipelines.}
Comparing \textsc{HELO-APR} with \textit{LANTERN} highlights the advantage of training-based cross-lingual knowledge transfer over multi-agent translation pipelines.
As shown in Table~\ref{tab:main_results}, \textsc{HELO-APR}  outperforms \textit{LANTERN} across most metrics, especially on the CodeLlama backbone.
This gap indicates that translation-based repair without targeted fine-tuning is insufficient for smaller open-source LLMs, as repair knowledge in the target language remains weak even after translation.
In contrast, \textsc{HELO-APR} internalizes cross-lingual repair knowledge through curriculum-driven fine-tuning, leading to more accurate and stable LRPL repair.
Meanwhile, although \textit{LANTERN} requires no additional training, its iterative pipeline of translation, repair, and back-translation introduces significantly higher inference cost.

\vspace{5pt}
\begin{tcolorbox}[title=\textbf{Answer to RQ1}, colback=gray!5, colframe=black!75, left=2pt, right=2pt, top=2pt, bottom=2pt]
(1) \textsc{HELO-APR} achieves the strongest overall repair performance, obtaining the best average \textit{Pass@k} results while maintaining high Compilation Rates (e.g., 97.83\% on Ruby with CodeLlama).

(2) \textsc{HELO-APR} mitigates cross-lingual syntactic interference by eliminating source-language leakage (CR$_S$ reduced to 0.00\%), in contrast to \textit{HRPLs-Only}.

(3) Leveraging HRPL buggy-fixed pairs yields stronger LRPL repair performance than \textit{Wong et al.} (which generates buggy-fixed pairs from scratch), demonstrating the potential of leveraging HRPL data to improve LRPL repair.

(4) Compared with the training-free multi-agent approach \textit{LANTERN}, \textsc{HELO-APR} achieves better overall repair performance while avoiding the high inference cost associated with iterative translation-based pipelines.

\end{tcolorbox}

\subsection{RQ2: LRPLs Dataset Construction}
\label{sec:rq2}

In this section, we investigate how different components of the proposed dataset construction pipeline contribute to the quality of the synthesized LRPL dataset.
Specifically, we first analyze the individual impact of each pipeline component through controlled ablation studies (RQ2.1), and then conduct a focused quality analysis to interpret the observed performance differences (RQ2.2).

\subsubsection{Contribution of Individual Pipeline Components (RQ2.1)}
\label{sec:rq2_1}

To analyze the contribution of each component in the dataset construction, we compare \textsc{HELO-APR} with four ablated variants, each removing or modifying a specific functional module.
Since dataset quality cannot be directly observed, we assess the contribution of each component indirectly through its downstream impact on repair performance under controlled training settings, following common practice in learning-based APR.

\vspace{0.2em}
\noindent\textbf{Impact of Transferability Analysis.}
\begin{itemize}[leftmargin=*] 
    \item \textit{w/o Transferability Analysis:} This variant removes the transferability analysis. Instead of excluding samples whose defects rely on non-transferable language features (e.g., pointer arithmetic), it translates all samples without filtering. This ablation evaluates whether transferability analysis is necessary.
\end{itemize}

\vspace{0.2em}
\noindent\textbf{Impact of Translation \& Injection Strategy.}
\textsc{HELO-APR} applies structural constraints only to defect-related regions. To verify this design, we compare this strategy with two extreme approaches: one with no constraints and one with global structural constraints.
\begin{itemize}[leftmargin=*]
    \item \textit{w/o Structure Constraints:} 
    This variant removes all structural constraints and simply prompts the LLM to translate the code and inject defects directly. 

    \item \textit{w/ Global Structure Constraints:} 
    This variant enforces structure constraints over the \textit{entire} program by conservatively assuming that all code regions may contain defect injection anchors. As a result, any rewriting that could remove or obscure such anchors is globally suppressed, even in defect-irrelevant regions.
\end{itemize}

\vspace{0.2em}
\noindent\textbf{Impact of Verification.}
\begin{itemize}[leftmargin=*]
    \item \textit{w/o Test Verification:} 
    This variant removes all verification steps, including \textit{Correctness Verification} and \textit{Defect Behavior Assessment}. This ablation evaluates the role of the verification phase in improving dataset quality.
\end{itemize}

\begin{table*}[t]
\centering
\caption{Ablation Study on Data Construction Strategies. We evaluate the impact of Transferability Analysis, Translation \& Injection Strategies, and Verification. }
\label{tab:ablation_study}
\renewcommand{\arraystretch}{1.2} 
\resizebox{\textwidth}{!}{%
\setlength{\tabcolsep}{3.5pt} 
\begin{tabular}{ll ccc @{\hspace{0.8em}} ccc @{\hspace{1.5em}} ccc @{\hspace{0.8em}} ccc}
\toprule

\multirow{3}{*}{\textbf{Strategy}} & \multirow{3}{*}{\textbf{Lang.}} 
& \multicolumn{6}{c}{\textbf{Backbone A: CodeLlama-7B}} 
& \multicolumn{6}{c}{\textbf{Backbone B: DeepSeek-Coder-6.7B}} \\
\cmidrule(lr){3-8} \cmidrule(lr){9-14}

& 
& \multicolumn{3}{c}{\textit{Dataset-Only}} & \multicolumn{3}{c}{\textit{Full-Pipeline}} 
& \multicolumn{3}{c}{\textit{Dataset-Only}} & \multicolumn{3}{c}{\textit{Full-Pipeline}} \\ 
\cmidrule(lr){3-5} \cmidrule(lr){6-8} \cmidrule(lr){9-11} \cmidrule(lr){12-14}

& 
& P@1 & P@3 & P@5 & P@1 & P@3 & P@5 
& P@1 & P@3 & P@5 & P@1 & P@3 & P@5 \\
\midrule

\multirow{2}{*}{\shortstack[l]{\textit{w/o Transferability}\\\textit{Analysis}}} 
& Ruby & 9.91 & 14.24 & 15.48 & 10.53 & 15.17 & 17.03 & 33.44 & 42.72 & 46.13  & 37.77  & 46.13  &  48.30 \\
& Rust & 6.36 & 11.56 & 12.72 & 7.51 & 12.14 & 13.29 & 44.51 & 50.87 & 52.60 & 45.66 & 51.45 & 54.91 \\
\midrule

\multirow{2}{*}{\shortstack[l]{\textit{w/o Structure}\\\textit{Constraints}}}
& Ruby & 8.67 & 12.38 & 14.24 & 10.84 & 16.72 & 18.27 & 32.82 & 40.87 & 45.20 & 38.08 & 47.68 & 49.23 \\
& Rust & 6.36 & 10.40 & 11.56 & 8.09 & 13.29 & 14.45 & 43.35 & 48.55 & 52.02 & 49.71 & 55.49 & 57.23 \\
\midrule

\multirow{2}{*}{\shortstack[l]{\textit{w/ Global Structure}\\\textit{Constraints}}}
& Ruby & 8.05 & 12.69 & 13.62 & 8.98 & 14.86 & 17.65 & 25.39 & 34.06 & 36.84 & 31.58 & 41.80 & 43.34 \\
& Rust & 5.78 & 8.67 & 9.83 & 9.83 & 9.83 & 10.40 & 32.95 & 41.04 & 45.66 & 46.24 & 51.45 & 52.02 \\
\midrule

\multirow{2}{*}{\shortstack[l]{\textit{w/o Test}\\\textit{Verification}}} 
& Ruby & 0.62 & 0.93 & 0.93 & 0.62 & 0.62 & 0.93 
& 4.64 & 5.57 & 5.88 & 4.95 & 5.57 & 5.88 \\
& Rust & 0.58 & 1.16 & 1.73 & 0.58 & 1.16 & 1.73 
& 2.89 & 3.47 & 4.05 & 2.89 & 4.05 & 4.62 \\
\midrule

\rowcolor{gray!10}
\textsc{HELO-APR} & Ruby & 
\textbf{10.22} & \textbf{15.17} & \textbf{16.72} & 
\textbf{12.38} & \textbf{17.34} & \textbf{19.50} & 
\textbf{36.53} & \textbf{44.89} & \textbf{47.99} & 
\textbf{41.80} & \textbf{50.15} & \textbf{53.56} \\

\rowcolor{gray!10}
 & Rust & 
\textbf{8.09} & \textbf{12.14} & \textbf{13.87} & 
\textbf{11.56} & \textbf{18.50} & \textbf{19.65} & 
\textbf{45.09} & \textbf{51.45} & \textbf{53.76} & 
\textbf{55.49} & \textbf{60.12} & \textbf{62.43} \\

\bottomrule
\end{tabular}%
}
\end{table*}

\textbf{Results and Analysis}:
Table~\ref{tab:ablation_study} presents the results of our ablation study. Performance is evaluated under two settings:(1) training exclusively on the synthesized data (\textit{Dataset-Only}), and (2) continuing to apply the Cross-Language Knowledge Transfer pipeline (\textit{Full-Pipeline}). The results confirm that each component of the proposed pipeline is essential.

\vspace{0.5em}
\noindent\textbf{Impact of Transferability Analysis.}
Table~\ref{tab:ablation_study} shows that the performance difference between the \textit{w/o Transferability Analysis} variant and \textsc{HELO-APR} is marginal in the \textit{Dataset-Only} setting.
For instance, on Rust with the DeepSeek backbone under the Dataset-Only setting, the baseline performs comparably to \textsc{HELO-APR} (Pass@1: 44.51\% vs.\ 45.09\%).
However, in the \textit{Full-Pipeline} setting, this difference becomes substantially larger.
Taking Rust as an example, \textsc{HELO-APR} clearly outperforms the baseline (Pass@1: 45.66\% vs.\ 55.49\%; Pass@5: 54.91\% vs.\ 62.43\%).
This indicates that non-transferable defects introduce semantic noise that has a limited impact when learning LRPL repair patterns in isolation.
However, during cross-lingual knowledge transfer, such noise interferes with the semantic alignment between HRPL-LRPL buggy-fixed pairs, significantly impairing alignment effectiveness.
Therefore, filtering non-transferable defects is a critical prerequisite for effective cross-lingual repair knowledge transfer.

\vspace{0.5em}
\noindent\textbf{Impact of Translation \& Injection Strategy.}
Table~\ref{tab:ablation_study} reveals a clear performance hierarchy among translation strategies.
\textsc{HELO-APR} consistently outperforms the \textit{w/o Structure Constraints} baseline, which also significantly outperforms the \textit{w/ Global Structure Constraints} variant.
For example, on the DeepSeek-Ruby task under the Full-Pipeline setting, \textsc{HELO-APR} achieves 41.80\% (Pass@1) and 53.56\% (Pass@5), compared to 38.08\% and 49.23\% for the unconstrained approach and 31.58\% and 43.34\% for global structural constraints.
These results indicate that selectively constraining defect-related regions is critical for synthesizing high-quality data, whereas enforcing global structural constraints is harmful.
We further analyze this phenomenon using representative Ruby examples in \S\ref{sec:rq2_2}.

\vspace{0.5em}
\noindent\textbf{Impact of Verification.}
As shown in Table~\ref{tab:ablation_study}, removing the verification phase results in a catastrophic degradation of repair performance across all languages and backbones.
For example, on CodeLlama-Ruby, Pass@1 drops to 0.62\% and Pass@5 to 0.93\%, while on Rust, the model becomes nearly ineffective (Pass@5: 1.73\%); even with the stronger DeepSeek backbone, all Pass@k scores remain below 6\%.
This collapse stems from two complementary factors: without \textit{Correctness Verification}, the synthesized training data frequently contains functionally invalid programs, causing the model to learn erroneous repair patterns; without \textit{Defect Behavior Assessment}, mismatches between synthesized LRPL defects and their HRPL counterparts break the semantic alignment required for effective cross-lingual knowledge transfer.

\begin{tcolorbox}[title=\textbf{Answer to RQ2.1}, colback=gray!5, colframe=black!75, left=2pt, right=2pt, top=2pt, bottom=2pt]
Each component of the LRPL dataset construction pipeline is critical to synthesizing high-quality training data:

(1) \textit{Transferability Analysis} filters out defects that cannot be reproduced in LRPLs. Without this step, non-transferable samples introduce semantic inconsistency between HRPL-LRPL pairs, significantly impairing cross-lingual knowledge transfer.

(2) \textit{Structure-Constrained Translation \& Injection} applies constraints only to defect-related regions, which consistently leads to better repair performance than both \textit{w/o Structure Constraints} and \textit{w/ Global Structure Constraints}.

(3) \textit{Test-Driven Verification} serves as a necessary quality-control mechanism. Removing this step leads to a substantial performance degradation, as verification is essential for ensuring both functional correctness and defect consistency.
\end{tcolorbox}

\subsubsection{Interpreting Dataset Quality: Semantic Fidelity vs. Code Naturalness (RQ2.2)}
\label{sec:rq2_2}

In \S\ref{sec:rq2_1}, we observed a clear performance hierarchy among different dataset construction strategies:
\textsc{HELO-APR} $>$ \textit{w/o Structure Constraints} $>$ \textit{w/ Global Structure Constraints}.
To further characterize how the quality of the synthesized LRPL datasets varies across different construction strategies, we conduct a focused analysis along two complementary dimensions: \textit{Defect Semantic Fidelity} and \textit{Code Naturalness}.

\vspace{0.5em}
\noindent\textbf{Metric 1: Defect Semantic Fidelity.}
This metric evaluates whether synthesized LRPL buggy-fixed pairs faithfully reproduce the defect semantics and execution behavior of their HRPL counterparts.
We randomly sampled 500 instances from the generated corpora.
Each sample was independently evaluated by three Ph.D. students proficient in Ruby using a 3-point ordinal scale:
\begin{itemize}[leftmargin=*]
    \item \textbf{Score 3 (High Fidelity):} The synthesized pair precisely reproduces the defect logic and trigger conditions of the HRPL source, with an identical root cause.
    \item \textbf{Score 2 (Partial Fidelity):} The synthesized pair preserves the defect type but deviates in trigger conditions or observable side effects.
    \item \textbf{Score 1 (No Fidelity):} The synthesized pair fails to reproduce the intended defect, e.g., by being functionally correct, syntactically invalid, or exhibiting an unrelated defect.
\end{itemize}
The final score for each sample was determined by majority agreement, and we report the average score for each strategy.

\vspace{0.5em}
\noindent\textbf{Metric 2: Code Naturalness.}
This metric measures the extent to which synthesized code adheres to idiomatic conventions of the target language.
We employ \textit{RuboCop}, a widely adopted static analyzer for Ruby, focusing on the \texttt{Style/*} category, which comprises over 80 rules capturing stylistic and idiomatic consistency.
We quantify code naturalness using \textit{Style Violation Density (SVD)}:
\begin{equation}
\small
\text{SVD} = \frac{\#(\text{Style Cop Violations})}{\text{Total LOC}} \times 1000
\end{equation}
Lower SVD values indicate closer adherence to community-adopted Ruby idioms.

\begin{table}[t]
\centering
\caption{Comparison of defect semantic fidelity and code naturalness across different dataset construction strategies.}
\label{tab:fidelity_style}
\small
\renewcommand{\arraystretch}{1}
\setlength{\tabcolsep}{4pt}
\begin{tabular}{lcc}
\toprule
\textbf{Strategy} & \textbf{Semantic Fidelity} $\uparrow$ & \textbf{Style Violation Density (SVD)} $\downarrow$ \\
\midrule
\textit{w/o Structure Constraints} & 2.24 & \textbf{68.84} \\
\textit{w/ Global Structure Constraints} & 2.64 & 89.52 \\
\rowcolor{gray!10} \textsc{HELO-APR} & \textbf{2.66} & 69.96 \\
\bottomrule
\end{tabular}
\end{table}

\vspace{0.5em}
\noindent\textbf{Results and Analysis.}
Table~\ref{tab:fidelity_style} summarizes the evaluation results.
Regarding \textit{Defect Semantic Fidelity}, both \textsc{HELO-APR} (2.66) and \textit{w/ Global Structure Constraints} (2.64) achieve substantially higher scores than the unconstrained strategy (2.24), indicating that introducing structural constraints helps preserve defect semantics during translation and defect injection.

In contrast, \textit{Code Naturalness} exhibits an opposite trend.
Applying structural constraints globally results in a markedly higher SVD (89.52), suggesting that strict global preservation of source structure leads to non-idiomatic, source-language-biased Ruby code.
By selectively constraining only defect-related regions, \textsc{HELO-APR} maintains a low violation density (69.96), comparable to the unconstrained baseline (68.84), while retaining high semantic fidelity.

\vspace{5pt}
\begin{tcolorbox}[title=\textbf{Answer to RQ2.2}, colback=gray!5, colframe=black!75, left=2pt, right=2pt, top=2pt, bottom=2pt]
The performance differences observed in RQ2.1 can be explained by a trade-off between defect semantic fidelity and code naturalness:

(1) \textit{Defect semantic fidelity.}
Introducing structural constraints helps preserve defect semantics during translation and defect injection.
Accordingly, both \textsc{HELO-APR} and globally constrained translation achieve substantially higher fidelity scores than the unconstrained strategy.

(2) \textit{Code naturalness.}
Applying structural constraints globally degrades code naturalness, producing non-idiomatic target-language code.
By contrast, selectively constraining only defect-related regions allows \textsc{HELO-APR} to maintain code naturalness at a level comparable to unconstrained translation.

(3) \textit{Overall trade-off.}
By balancing high semantic fidelity with high code naturalness, \textsc{HELO-APR} achieves the most favorable data quality, which explains its superior downstream repair performance.
\end{tcolorbox}

\subsection{RQ3: Impact of Cross-Language Knowledge Transfer}
\label{sec:rq3}

\begin{table*}[t]
\centering
\caption{Ablation study on different training stages. M2 examines the effect of removing HRPL Repair Learning (Stage~1), while M3 investigates the impact of removing Cross-Lingual Repair Alignment (Stage~2). M4 corresponds to the full \textsc{HELO-APR} curriculum.}
\label{tab:curriculum_ablation}
\resizebox{\textwidth}{!}{%
\begin{tabular}{l ccc ccc ccc}
\toprule
\multirow{2}{*}{\textbf{Method}} & 
\multicolumn{3}{c}{\textbf{Ruby}} & 
\multicolumn{3}{c}{\textbf{Rust}} & 
\multicolumn{3}{c}{\textbf{Average}} \\
\cmidrule(lr){2-4} \cmidrule(lr){5-7} \cmidrule(lr){8-10}
 & \textbf{P@1 (\%)} & \textbf{P@3 (\%)} & \textbf{P@5 (\%)} & \textbf{P@1 (\%)} & \textbf{P@3 (\%)} & \textbf{P@5 (\%)} & \textbf{P@1 (\%)} & \textbf{P@3 (\%)} & \textbf{P@5 (\%)} \\
\midrule

\multicolumn{10}{l}{\textit{\textbf{Backbone A: CodeLlama-7B}}} \\
\midrule
M1: Direct Finetuning (S3 Only)                    & 10.22 & 15.17 & 16.72 & 8.09  & 12.14 & 13.87 & 9.16 & 13.66 & 15.30 \\
M2: w/o HRPL Repair Learning (S2 $\rightarrow$ S3) & 10.84 & 16.41 & 17.03 & 9.83  & 15.61 & 16.18 & 10.34 & 16.01 & 16.61 \\
M3: w/o Cross-Lingual Repair Alignment (S1 $\rightarrow$ S3)   & 11.46 & 15.79 & 17.03 & 9.83  & 16.76 & 17.92 & 10.65 & 16.28 & 17.48 \\
\rowcolor{lightgray} 
M4: \textsc{HELO-APR} (Full stages)                & \textbf{12.38} & \textbf{17.34} & \textbf{19.50} & \textbf{11.56} & \textbf{18.50} & \textbf{19.65} & \textbf{11.97} & \textbf{17.92} & \textbf{19.58} \\

\midrule
\addlinespace 

\multicolumn{10}{l}{\textit{\textbf{Backbone B: DeepSeek-Coder-6.7B}}} \\
\midrule
M1: Direct Finetuning (S3 Only)                    & 36.53 & 44.89 & 47.99 & 45.09 & 51.45 & 53.76 & 40.81 & 48.17 & 50.88 \\
M2: w/o HRPL Repair Learning (S2 $\rightarrow$ S3) & 38.39 & 47.06 & 50.77 & 48.55 & 54.91 & 56.07 & 43.47 & 50.99 & 53.42 \\
M3: w/o Cross-Lingual Repair Alignment (S1 $\rightarrow$ S3)   & 37.46 & 46.44 & 49.85 & 49.71 & 54.91 & 56.65 & 43.59 & 50.68 & 53.25 \\
\rowcolor{lightgray} 
M4: \textsc{HELO-APR} (Full stages)                & \textbf{41.80} & \textbf{50.15} & \textbf{53.56} & \textbf{55.49} & \textbf{60.12} & \textbf{62.43} & \textbf{48.65} & \textbf{55.14} & \textbf{58.00} \\
\bottomrule
\end{tabular}
}
\end{table*}

To investigate the impact of cross-language knowledge transfer and assess the necessity of each training stage in \textsc{HELO-APR}, we conduct a comprehensive ablation study.
We design four experimental settings (M1-M4) that progressively remove individual stages from the proposed three-stage curriculum, disentangling the effects of \textit{HRPL Repair Learning} (Stage~1), \textit{Cross-Lingual Repair Alignment} (Stage~2), and \textit{LRPL Adaptation} (Stage~3).

\textit{Ablation Settings.}
\begin{itemize}[leftmargin=*]
    \item \textit{M1: Target-Only Baseline (S3 Only).}
    The model is fine-tuned exclusively on the synthesized LRPL dataset ($\mathcal{D}_{tgt}$), without any prior cross-language knowledge transfer.
    
    \item \textit{M2: w/o HRPL Repair Learning (S2 $\rightarrow$ S3).}
    This variant skips HRPL repair learning (Stage~1) and starts directly from cross-lingual repair alignment using parallel data ($\mathcal{D}_{para}$), followed by LRPL adaptation.
    
    \item \textit{M3: w/o Cross-Lingual Repair Alignment (S1 $\rightarrow$ S3).}
    This variant performs sequential fine-tuning on HRPL data ($\mathcal{D}_{src}$) and LRPL data ($\mathcal{D}_{tgt}$), omitting the explicit alignment stage.
    
    \item \textit{M4: \textsc{HELO-APR} (Full Pipeline).}
    This setting employs the complete three-stage curriculum (S1 $\rightarrow$ S2 $\rightarrow$ S3) to fully enable cross-language knowledge transfer.
\end{itemize}

To ensure a fair comparison across M1-M4, we adopt an early-stopping-based convergence criterion instead of a fixed training schedule.
This design prevents bias caused by under- or over-training and ensures that observed performance differences primarily reflect the effectiveness of different curriculum strategies.

The quantitative results are summarized in Table~\ref{tab:curriculum_ablation}.

\vspace{1ex}
\noindent\textbf{Overall Impact of Cross-Language Knowledge Transfer.}
As shown in Table~\ref{tab:curriculum_ablation}, the full model \textit{M4 (\textsc{HELO-APR})} consistently achieves the best performance across all backbones and evaluation metrics.
For example, with the DeepSeek-Coder-6.7B backbone, M4 attains an average Pass@1 of 48.65\%, substantially outperforming the target-only baseline M1 (40.81\%).
This consistent improvement across different $k$ values indicates that the proposed curriculum effectively enhances the model’s ability to retrieve correct fixes under limited and extended candidate budgets.

\vspace{1ex}
\noindent\textbf{Effect of HRPL Repair Learning (Stage~1).}
Comparing \textbf{M2} with the full model \textbf{M4}, we observe a consistent performance degradation when HRPL repair learning is removed.
On DeepSeek-Coder-6.7B, omitting Stage~1 results in a drop of 5.18\% in Pass@1 and a noticeable reduction in Pass@5.
Similar trends are observed with the CodeLlama backbone.
These results indicate that HRPL repair learning provides a necessary foundation of general repair capabilities; without this stage, the model lacks sufficient repair knowledge to fully benefit from subsequent training.

\vspace{1ex}
\noindent\textbf{Effect of Cross-Lingual Repair Alignment (Stage~2).}
The results for \textbf{M3} further highlight the critical role of explicit cross-lingual alignment.
Despite being trained on abundant HRPL data, models without Stage~2 exhibit a substantial performance drop.
For instance, removing this stage reduces Pass@1 by 5.06\% on DeepSeek-Coder-6.7B.
This finding suggests that HRPL repair knowledge alone is insufficient for effective LRPL repair.
By explicitly aligning repair behaviors across languages using parallel buggy-fixed pairs, Stage~2 enables the model to transfer learned repair capabilities into the target language, thereby substantially improving repair performance.

\vspace{5pt}
\begin{tcolorbox}[title=\textbf{Answer to RQ3}, colback=gray!5, colframe=black!75, left=2pt, right=2pt, top=2pt, bottom=2pt]
Our ablation study demonstrates that effective cross-language knowledge transfer requires a carefully designed multi-stage curriculum.

(1) \textit{HRPL Repair Learning (Stage~1)} provides a necessary foundation of general repair knowledge but is not sufficient on its own for effective LRPL repair.

(2) \textit{Cross-Lingual Repair Alignment (Stage~2)} plays a pivotal role by explicitly aligning repair behaviors across languages, enabling HRPL repair knowledge to be effectively realized in LRPLs.

(3) Combining both stages with LRPL adaptation yields the best performance across all backbones and metrics, confirming that cross-language knowledge transfer is maximized only under the full curriculum.
\end{tcolorbox}

\subsection{RQ4: Generalizability to Real-World Development Scenarios}
\label{sec:rq4}

To evaluate whether repair knowledge learned through cross-lingual training generalizes beyond synthetic benchmarks, we assess \textsc{HELO-APR} on real-world Ruby bugs from Defects4Ruby. Unlike xCodeEval, this benchmark consists of bugs written and fixed by human developers, reflecting realistic development scenarios.

We randomly sample 1000 buggy-fixed pairs for evaluation and compare the base model against the same backbone fine-tuned with \textsc{HELO-APR}.
Following common practice on Defects4Ruby, we adopt text-based similarity metrics, including BLEU-4 and ROUGE-1, to measure the overlap between generated patches and developer-written fixes.
All experiments are conducted with two backbones: CodeLlama-7B and DeepSeek-Coder-6.7B.

\begin{table*}[t]
\centering
\caption{Generalization results on a randomly sampled subset (1000 instances) of Defects4Ruby.
We report \textit{BLEU-4} and \textit{ROUGE-1} for real-world Ruby program repair.
Best results for each metric and backbone are highlighted in bold.}
\label{tab:rq4_defects4ruby}
\renewcommand{\arraystretch}{1.15}
\setlength{\tabcolsep}{5pt} 
\begin{footnotesize}
\begin{tabular}{l cc @{\hspace{1em}} cc}
\toprule
\multirow{2}{*}{\textbf{Method}} 
& \multicolumn{2}{c}{\textbf{CodeLlama-7B}} 
& \multicolumn{2}{c}{\textbf{DeepSeek-Coder-6.7B}} \\
\cmidrule(lr){2-3} \cmidrule(lr){4-5}
& BLEU-4 & ROUGE-1 
& BLEU-4 & ROUGE-1 \\
\midrule
Zero Shot       
& 61.20 & 76.76 
& 58.13 & 71.70 \\

\rowcolor{gray!10}
\textsc{HELO-APR} 
& \textbf{66.79} & \textbf{83.59} 
& \textbf{73.93} & \textbf{87.60} \\

\bottomrule
\end{tabular}
\end{footnotesize}
\end{table*}

As shown in Table~\ref{tab:rq4_defects4ruby}, \textsc{HELO-APR} consistently outperforms the Zero-Shot baseline on Defects4Ruby across both backbones.
In particular, it improves BLEU-4 and ROUGE-1 on CodeLlama-7B (61.20→66.79, 76.76→83.59) and achieves larger gains on DeepSeek-Coder-6.7B (58.13→73.93, 71.70→87.60).

These results indicate that patches generated by HELO-APR exhibit higher textual overlap
and stylistic proximity to developer-written fixes. Although text-based similarity metrics do not
directly measure functional correctness, the consistent improvements provide supportive external
evidence that cross-lingual repair knowledge leads to outputs better aligned with the textual and
stylistic characteristics of real-world patches.

\begin{tcolorbox}[title=\textbf{Answer to RQ4}, colback=gray!5, colframe=black!75, left=2pt, right=2pt, top=2pt, bottom=2pt]
On Defects4Ruby, HELO-APR consistently outperforms the Zero-Shot baseline on text-based similarity metrics across both backbones.
This provides supportive external evidence that cross-lingual repair knowledge improves the alignment between generated patches and developer-written fixes in real-world settings.
\end{tcolorbox}

\section{Threats to Validity}

\textit{Internal Validity.}
Internal validity concerns whether the observed improvements can be attributed to \textsc{HELO-APR} rather than experimental confounders.
Different curriculum variants may converge at different rates; under a fixed training schedule, some variants may underfit while others overfit, potentially biasing performance comparisons.
To mitigate this risk, we adopt an early-stopping-based convergence criterion for all variants, ensuring evaluation near their respective convergence points and reducing the influence of training-time differences.

In addition, the results may be influenced by the choice of model scale as part of the experimental setting. Our experiments are conducted on 7B-scale models, and the effectiveness of \textsc{HELO-APR} on larger models has not yet been validated, which may limit the generalizability of our findings. Nevertheless, 7B models are representative in efficiency-sensitive and deployment-constrained scenarios, and thus our results still carry practical significance.

\textit{External Validity.}
External validity concerns the extent to which our findings extend beyond \textsc{xCodeEval}, whose competition-based data distribution differs from real-world software development.
To provide external evidence beyond synthetic benchmarks, we directly evaluate models trained on \textsc{xCodeEval} on \textsc{Defects4Ruby} without additional fine-tuning.
Although this evaluation relies on text-based similarity metrics rather than functional correctness, the consistent improvements indicate that \textsc{HELO-APR} produces patches more closely aligned with developer-written fixes in real-world settings.

\section{Related Work}
\label{sec:background}

\subsection{LLM-Based Program Repair}
\label{sec:bg_apr}

LLMs have significantly advanced APR, leading to successive paradigm shifts~\cite{jiang2021cure,yang2024swe,zhang2024appt,yuan2024designrepair}.
Early work formulated APR as conditional sequence generation using \emph{NMT-style} or \emph{clozing-style} models~\cite{hua2018towards}, exemplified by TFix~\cite{berabi2021tfix}, VulRepair~\cite{fu2022vulrepair}, and AlphaRepair~\cite{xia2022less}.
More recent approaches adopt \emph{conversational repair}, enabling iterative reasoning over bugs and patches via natural language interaction~\cite{xia2024automated,kong2024contrastrepair}.
Building on this trend, \emph{agentic APR} further leverages multiple LLM agents to collaboratively perform localization, patch generation, and validation~\cite{antoniades2025swesearchenhancingsoftwareagents,bouzenia2024repairagentautonomousllmbasedagent,ye2025adversarial,ma2025alibaba,zhang2024autocoderover,yang2025survey}.

Despite these advances, LLM-driven APR research has largely focused on high-resource programming languages~\cite{zhang2023survey}.
For low-resource programming languages (LRPLs) such as Kotlin, Ruby, and PHP, APR remains underexplored, and the scarcity of high-quality parallel buggy-fixed corpora further limits the applicability of state-of-the-art LLM-based repair techniques~\cite{zhang2025bridge,joel2024survey}.

\subsection{Program Repair for LRPLs}
\label{sec:bg_transfer}

APR for LRPLs is mainly constrained by the scarcity of high-quality buggy–fixed corpora~\cite{zhang2023survey,luo2025unlocking}.
Existing work addresses this limitation from several directions.
Some approaches transfer repair knowledge by directly fine-tuning on HRPL corpora or multilingual code models, but often suffer from syntactic interference~\cite{baltajicross}.
Others leverage LLMs to synthesize buggy–fixed pairs for data augmentation~\cite{wong2025investigating}, though the generated defects are frequently shallow or unreliable.
Related work combines cross-language code translation with downstream repair to reuse HRPL repair capabilities, but this pipeline heavily depends on translation quality and is less effective for relatively small, open-source LLMs~\cite{luo2025unlocking}.
In addition, meta-learning-based methods support few-shot adaptation to LRPL defects, yet still require a non-trivial amount of labeled LRPL data~\cite{wang2023towards}.

In contrast, \textsc{HELO-APR} transfers repair knowledge from HRPLs to LRPLs in a more reliable manner.
It constructs cross-lingual parallel buggy–fixed pairs via a principled LRPL data generation strategy and employs a curriculum learning–based cross-lingual knowledge transfer pipeline with an explicit repair alignment stage, enabling more robust knowledge transfer to LRPLs.

\section{Conclusion}
This paper presents \emph{HELO-APR}, a two-stage cross-lingual supervision framework that converts abundant HRPL repair signals into verified LRPL training data and transfers them via curriculum fine-tuning.
On xCodeEval with C++ as the source HRPL and Ruby and Rust as the target LRPLs, \emph{HELO-APR} improves average Pass@1 from 31.17\% to 48.65\% on DeepSeek-Coder-6.7B and from 1.67\% to 11.97\% on CodeLlama-7B.
It also strengthens syntactic validity by raising the average target compilation rate on CodeLlama from 49.77\% to 91.98\%, reflecting substantially reduced cross-lingual syntactic interference.
On Defects4Ruby, HELO-APR produces patches with higher textual similarity to developer-written fixes, increasing BLEU-4 from 61.20 to 66.79 and ROUGE-1 from 76.76 to 83.59 on CodeLlama-7B.
These results suggest that verified cross-lingual supervision can serve as a scalable recipe for unlocking LLM repair capabilities in low-resource settings.
The same principle may extend beyond APR to other code tasks where target-language supervision is scarce, by synthesizing and validating task-specific cross-lingual training signals and explicitly aligning behaviors across languages.

\bibliographystyle{ACM-Reference-Format}
\bibliography{arxiv}

\end{document}